\documentclass[11pt]{article}

\usepackage{hyperref}
\usepackage{url}
\usepackage{graphicx}
\usepackage{array}

\begin{document}


\begin{center}
\vbox to 1 truecm {}
{\Large \bf Convoluting jet spectra with fragmentation functions:} \\[0.2cm]
{\Large \bf a cross-check of the charged particle $p_T$ spectrum} \\[0.8cm]
{\large Andre S. Yoon, Edward Wenger, and Gunther Roland }\\[0.2cm]
{\it  Laboratory for Nuclear Science, MIT, Cambridge, MA 02139-4307, USA}\\[1.4cm]
\vskip 5 truemm
\end{center}


\begin{abstract}
Motivated by the excellent agreement between next-to-leading-order pQCD calculations and the inclusive jet spectra measured by CDF, we cross-check PYTHIA fragmentation functions. 
The convolution of the measured jet spectra with unmodified PYTHIA fragmentation functions results in reasonable agreement with the PYTHIA charged particle spectrum over the entire $p_T$ range of interest, while there is a sizable disagreement with the measured charged particle spectrum above $p_T=30$~GeV/c. 
In an attempt to understand the source of this discrepancy, we introduce a number of increasingly different toy-model fragmentation functions for the convolution.
However, even the most extreme fragmentation functions result in an underestimate of the high-$p_T$ CDF spectrum, which remains irreconcilable with the measured jet spectra.
 
\end{abstract}
\newpage

\section{Introduction}

After an intense period of commissioning and a short period of collisions at 0.9 and 2.36 TeV
in 2009~\cite{LHCcom}, the continuous operation of the Large Hadron Collider (LHC) at multi-TeV collision energies looms just around the corner. 
Among the first measurements at the LHC will be the inclusive production of single charged particles (or hadrons), i.e. $pp \rightarrow h+X$, measured differentially in pseudorapidity ($\eta$) and transverse momentum ($p_T$)~\cite{EarlyAtLHC}. 
While both distributions are the subject of Quantum Chromodynamics (QCD), 
the former is generally modeled phenomenologically due to the non-perturbative nature of low-$p_T$ bulk production, while the latter, which involves hard processes, is generally described by the perturbative theory of QCD (pQCD). 
Hard production of high transverse momenta particles ($p_T \geq 2$ GeV/c) originates from the fragmentation of hard-scattered partons~\cite{QCDFactorized}.
The fragmentation of hard-scattered partons into hadrons is described by the probability of finding a hadron carrying a specific fraction of the parton momentum, known as the fragmentation function (FF). 
In hadronic collisions, a full description also requires knowledge of the distribution of the initial partons within the colliding hadrons, known as the parton distribution function (PDF). 
The measurement of the inclusive charged particle $p_T$ spectrum at large transverse momentum, therefore, measures in essence the convolution of three pieces: the hard-parton scattering cross section, the PDFs and the FFs.

Especially at LHC energies, where a large fraction of the total cross section is comprised of the underlying QCD dynamics, a precise understanding of the QCD background rates is not only important for understanding Standard Model particle production ($W^{\pm}$, Z, Higgs), but also for rare processes beyond the Standard Model~\cite{LHCqcd,LHCqcd2}.
In addition, the inclusive charged particle $p_T$ spectrum in $pp$ collisions is an important reference for studying high-$p_T$ particle suppression in the dense QCD medium produced in high energy nucleus-nucleus (AA) collisions~\cite{JetQ,QGP}. 
The suppression (or enhancement) of high-$p_T$ particles is typically quantified by the ratio of charged particle $p_T$ spectra in $AA$ collisions to those in $pp$ collisions scaled by the number of binary nucleon-nucleon collisions, known as the nuclear modification factor $R_{AA}$~\cite{QGP}.  
At RHIC, the factor of 5 suppression seen in $R_{AA}$~\cite{RaaBR,RaaPH,RaaPHOBOS,RaaST} was an early indication of strong final-state medium effects on particle production.
It is similarly expected to be one of the first measurements performed by the heavy ion programs at the LHC~\cite{CMShi}. 
 
Experimentally, the inclusive charged particle $p_T$ spectra have been measured in $pp$ and $p\bar{p}$ over a wide range of center-of-mass energies from 31 GeV to 1.96 TeV~\cite{ISR_hipt,RHIC_hipt,UA1,CDFsingleE,CDFsingle}, and recently at 2.36 TeV at LHC~\cite{fCMS,fATLAS}. While the measurements up to  $\sqrt{s}=$1.8 TeV (and  $\sqrt{s}=$2.36 TeV) are limited to $p_{T}<20$ GeV/c, the latest CDF measurement~\cite{CDFsingle} at $\sqrt{s}=$1.96 TeV, based on an integrated luminosity of 506 $pb^{-1}$, extended the $p_T$ reach up to about 140 GeV/c for the first time. However, the measurement shows that the high-$p_T$ region cannot be described by the power-law modeling established in their earlier measurement at $\sqrt{s}$=1.8 TeV~\cite{CDFsingleE}. 
The incompatibility between the new measurement and the power-law modeling grows from 50$\%$ at $p_{T}=20$~GeV/c up to a factor of 1000 at $p_{T}=125$~GeV/c. 
In order to fit the entire $p_{T}$ range, a more sophisticated parameterization was introduced~\cite{CDFsingleE}, namely another power law term was added to the previous fit.  
With the new parameterization, the normalized chi-square ($\chi^{2}/ndf$) is reduced from 258/182 to 80/223, albeit with a factor of 5-7 discrepancy remaining above $\sim90$~GeV/c.
The similarity between the spectra observed at both collision energies up to $p_T=9$~GeV/c (the range measured at 1.8 TeV) suggests that the incompatibility cannot be accounted for by the 9$\%$ increase in center-of-mass energy. 

A disagreement of similar magnitude is observed when the measured high-$p_T$ spectra are compared to leading-order (LO) and next-to-leading-order (NLO) pQCD calculations, as well as a simple extrapolation based on $x_{T}$ scaling~\cite{FA_DD_AY,A_K_K}. 
The validity of the factorization theorem is even questioned in~\cite{A_K_K}, as a response to the huge discrepancy seen at high $p_{T}$ -- the regime where their NLO pQCD calculation should be most reliable. 
While the CDF paper~\cite{CDFsingle} offers no possible physics origin for the exceptionally large measured cross section at high $p_{T}$, the sizable incompatibility with not only their former power-law modeling but also (N)LO pQCD calculations and $x_{T}$ scaling suggests that further study might be necessary. 

At the LHC, the nominal heavy ion collisions ($PbPb$) will take place at a center-of-mass energy of 5.5 TeV per nucleon pair, corresponding to the nominal $pp$ collision energy of 14 TeV for the same magnetic rigidity \footnote{For the same magnetic rigidity of the LHC machine, the center-of-mass energy per nucleon pair in heavy ion collisions is just defined as the center-of-mass energy in $pp$ scaled by the charge-to-mass ratio of the lead ion: 82/208.}. For the first year, however, as the center-of-mass energy of $pp$ collisions will be limited to 7 TeV~\cite{LHCcom}, the corresponding $PbPb$ center-of-mass energy will be limited to 2.76 TeV per nucleon pair. 
Since the $R_{AA}$ measurement requires a $pp$ reference at the same collision energy as $PbPb$, the first-year heavy ion measurements at 2.76 TeV will rely on a combination of theory predictions and interpolations between lower energy measurements and those performed at 7 TeV. 
In this perspective, the CDF measurement is unique in two regards.  
First, the center-of-mass energy of 1.96 TeV is closest to that planned for the first-year heavy ion run (except for the measurement at 2.36 TeV which is limited in $p_T$ reach).  
Second, the reach to high $p_T$ far exceeds any previous measurements. 
Thus, understanding the observed discrepancy with the pQCD prediction is crucial. 

In contrast to the observed discrepancy in the charged hadron spectra, the CDF inclusive jet spectrum are in fact well described by NLO pQCD calculations~\cite{CDFjet1,CDFjet2,CDFjet3,CDFjet4}. 
This is of particular interest, as high-$p_{T}$ charged particles are understood in pQCD to be predominantly~\footnote{There is a contribution from the leptonic decays of weak gauge bosons, but this sub-leading processes is negligible in the inclusive spectrum. In PYTHIA it amounts to only 10$\%$ at most~\cite{FA_DD_AY}.  It is also possible for high-$p_{T}$ hadrons to originate in the absence of a jet in a higher-twist (HT) picture.  However, as the production of hadrons in HT is power-law-suppressed in $p_T$, that contribution should be negligible as well~\cite{HTArleo}.} the fragmentation products of hard-scattered partons from the collision, i.e. ``jets''. 
In fact, NLO pQCD calculations carried out for charged particle spectra differ only from those for jet spectra by the addition of a parameterization of jet fragmentation. 
It is shown in \cite{FA_DD_AY} that the uncertainties related to different parameterizations of fragmentation functions (FF) and parton distribution functions (PDF) only amount to 10$\%$ and 25$\%$, respectively. 
In this case, the only possible explanations for the exceptionally large measured cross section are either that the current modeling of fragmentation is dramatically incorrect (by a factor of 1000!), or there is a flaw in the measurement, or there is a breakdown in the QCD factorization theorem as suggested in~\cite{A_K_K}. 

In this paper, we attempt to reconcile the apparent discrepancy in the CDF inclusive charged particle spectrum by convoluting the CDF inclusive jet spectra measurement with a set of increasingly different fragmentation functions.
We start with the PYTHIA fragmentation functions to see what the CDF jet spectra imply for the charged hadron spectra absent any surprises in the fragmentation.
Then, we modify the fragmentation functions arbitrarily within the bounds of energy and momentum conservation to see if the measured hadron spectrum can be recreated.
Finally, by using the hardest imaginable fragmentation function -- each jet fragments into a single charged hadron -- we rule out the possibility that unexpectedly hard fragmentation is responsible for the exceptionally large measured cross section.

The rest of the paper is organized as follows. 
In section 2, we introduce the technical details of the convolution method. 
We then describe the measured inclusive jet spectra and the PYTHIA inclusive jet spectra in section 3. 
In section 4, we argue that the convolution method can effectively reproduce the charged particle spectra within PYTHIA, i.e. by convoluting the  PYTHIA jet spectrum with the PYTHIA fragmentation functions.  
We then show that the CDF jet spectra convoluted with the same PYTHIA fragmentation functions gives a similar result. 
In section 5, we compare to the CDF charged hadron spectrum the results of convoluting the CDF jet spectra with an arbitrary set of increasingly unrealistic fragmentation functions. 
The final section contains our conclusions and a further discussion on how this convolution method may be used to cross-check charged particle cross sections.

\section{Convolution Method}
\label{sec:method}

In the QCD factorization scheme of hadron-hadron collisions~\cite{QCDFactorized,QCDFactorized2}, the invariant cross section for inclusive high-$p_{T}$ hadron production is given by:

\begin{eqnarray}
\lefteqn{E_{\scriptscriptstyle{C}}\frac{d^{3}\sigma(AB\rightarrow CX)}{d^{3}p_{\scriptscriptstyle{C}}} = } \nonumber \\
& & \frac{1}{\pi} \; \sum\int_{0}^{1}dx_{a}\int_{0}^{1}dx_{b} \; q_{a}^{\scriptscriptstyle{A}}(x_{a};Q^{2})\,q_{b}^{\scriptscriptstyle{B}}(x_{b};Q^{2}) \; \frac{1}{z} \; D^{c}_{\scriptscriptstyle{C}}(z;Q^{2}) \; \frac{d\hat{\sigma}(ab\rightarrow cd)}{d\hat{t}},
\label{eq1}
\end{eqnarray}

\noindent where the parton distribution function $q_{a}^{\scriptscriptstyle{A}}(x_{a})$ describes the number density of constituents $a$ within hadron $A$ with longitudinal momentum fraction $x_{a}$ (in the range $x_{a} \rightarrow x_{a}+dx_{a}$).
The fragmentation function $D^{c}_{\scriptscriptstyle{C}}(z)$ represents the probability that parton $c$ hadronizes into $C$ carrying a fraction $z$ of the parton energy.  
$Q^{2}$ is the characteristic energy scale of the hard scattering. 
The LO cross section for the hard scattering of partons $a$ and $b$ at short distance is denoted by $\hat{\sigma}$. 
The summation is over all partons $a$, $b$, $c$, and $d$.
The hadronization of parton $d$ is implicit in the summation. 
A direct calculation of Eq.~\ref{eq1} is possible up to a certain order in $\alpha_{s}$ provided that $q_{a}^{\scriptscriptstyle{A}}(x_{a})$  and $D^{c}_{\scriptscriptstyle{C}}(z)$ are given. 

Similarly, the related cross section for inclusive jet production (\mbox{$AB\rightarrow Jet+X$}) can be given by the same equation, but with the term \mbox{$(1/z)\,D^{c}_{\scriptscriptstyle{C}}(z;Q^{2})$} replaced by \mbox{$\delta(1-z)$}, since \mbox{$\sum_{\scriptscriptstyle{C}}D^{c}_{\scriptscriptstyle{C}}(z) = \delta(1-z)$}.
The convolution method is essentially equivalent to evaluating Eq.~\ref{eq1} but based on measurements of the inclusive jet cross section and the fragmentation functions, provided that both are measured: 

\begin{eqnarray}
\sigma_{had} & = & \mbox{$PDF_{a/A}$} \;\; \otimes \;\; \mbox{$PDF_{b/B}$}  \;\; \otimes \;\;  \hat{\sigma}( \mbox{\it{hard parton scattering}}) \;\; \otimes \;\; \mbox{FF} \\ & = & \sigma_{jet} \;\; \otimes \;\; \mbox{FF}. \nonumber
\end{eqnarray}


\noindent In this case, it is clear that a knowledge of the inclusive jet cross section and the fragmentation functions associated by jet-$p_{T}$ are enough to reproduce the hadron spectra.
The convolution of the jet cross section weighted in each jet-$p_{T}$ bin by the associated fragmentation function can be cast into the following simple differential form:

\begin{equation}
\frac{d\sigma_{had}}{dp_{T}dy_{had}} \simeq \sum_{i = 0} \frac{d\sigma_{jet}}{dp_{T, jet}dy_{jet}}\vert_{p_{T,jet} = p^{i}_{T,jet}} \times \Delta p^{i}_{T,jet} \Delta y_{jet}\times F\!F_{i}(p_{T},p^{i}_{T,jet}),
\label{eq2_1}
\end{equation}

\noindent where the summation is over all jet-$p_{T}$ bins, and the custom-built fragmentation function $F\!F_{i}$ is defined as:

\begin{equation}
F\!F_{i}(p_{T},p^{i}_{T,jet}) \equiv \frac{\Delta (\frac{d\sigma_{had}}{dp_{T}dy_{had}})}{\Delta \sigma_{jet}}\Bigg\vert_{p_{T,jet} = p^{i}_{T,jet}}.
\label{eq3}
\end{equation}

\noindent The quantity $F\!F_{i}$ is just the transverse momentum differential cross section of charged particles per jet cross section at a certain jet-$p_{T}$ with finite bin size. 


The first term in Eq.~\ref{eq2_1} is well known not only from theoretical calculations but also from measurements of the inclusive jet cross section.
However, the fragmentation functions are not known from any measurement in the exact form that they are needed (i.e. Eq.~\ref{eq3}).
In particular, the convolution requires that $F\!F_{i}$ be measured in the same bins of jet-$p_T$ as the inclusive jet cross section measurement.

\section{Inclusive Jet Spectra (Data vs PYTHIA)}
\label{sec:jetspectra}

The inclusive jet cross sections have been reported on numerous occasions by the CDF collaboration, showing good agreement with NLO pQCD predictions for different jet algorithms\footnote{$k_T$~\cite{KTalgo} and midpoint cone~\cite{MidPalgo}  algorithms are used.} ~\cite{CDFjet1,CDFjet2,CDFjet3,CDFjet4}. 
In particular, their latest measurement of the jet cross section~\cite{CDFjet1}, which used a mid-point cone algorithm on 1.13 $fb^{-1}$ of $p\bar{p}$ collision data, agrees over a large range of jet-$p_{T}$ and rapidity not only with the NLO pQCD prediction within the respective experimental and theoretical uncertainties, but also with the previous CDF measurements using different jet-finding algorithms.  Good agreement with NLO pQCD predictions is also seen by the D$\emptyset$ measurement over a similar jet-$p_{T}$ range~\cite{D0Jet}.  While a direct comparison of the inclusive jet measurements between the two experiments is not possible due to different rapidity binning, systematic data-theory comparisons for the determination of LO, NLO, and NNLO pQCD PDFs show no significant inconsistency between the measurements~\cite{MSTW08}.
In order to test whether the disagreement~\cite{CDFsingle,FA_DD_AY} in the charged particle spectra between CDF and PYTHIA might originate from a difference in the respective jet cross sections, we compare the latest inclusive jet cross section measurement with that from PYTHIA.

To maintain similar statistics over a large range of jet-$p_{T}$, several QCD jets samples 
\footnote{(MSEL = 1) with $low$ $p_T$ process (ISUB = 95) added for $\hat{p_{T}}\rightarrow 0$ to avoid a divergent jet cross section.} 
were generated in bins of the hard parton momentum transfer ($\hat{p_{T}}$ = 0-15, 15-20, 20-30 GeV/c, etc.) using the  D6T~\cite{tuneD6T} tune
of PYTHIA 6.41 (or 6.42)~\cite{bPythia}.
The different $\hat{p_{T}}$ bins were then combined after properly weighting each sample by its corresponding cross section 
\footnote{The cross section for the first $\hat{p_T}$ bin was obtained by taking the difference of the minimum bias cross section with the sum of the cross sections for all other $\hat{p_T}$ bins.}. 
Several different jet finding algorithms
\footnote{Iterative cone, $k_T$, and SIS cone were used with a cone radius $R\equiv\sqrt{\Delta\phi^{2}+\Delta y^{2}}=0.7$} 
were used at MC level to test for possible algorithmic dependences of the jet cross section.
The variation was observed to be smaller than the uncertainties in the measurement over the entire jet-$p_T$ range. 

\begin{figure}[htbp]
	\centering
    \includegraphics[width=7.7cm]{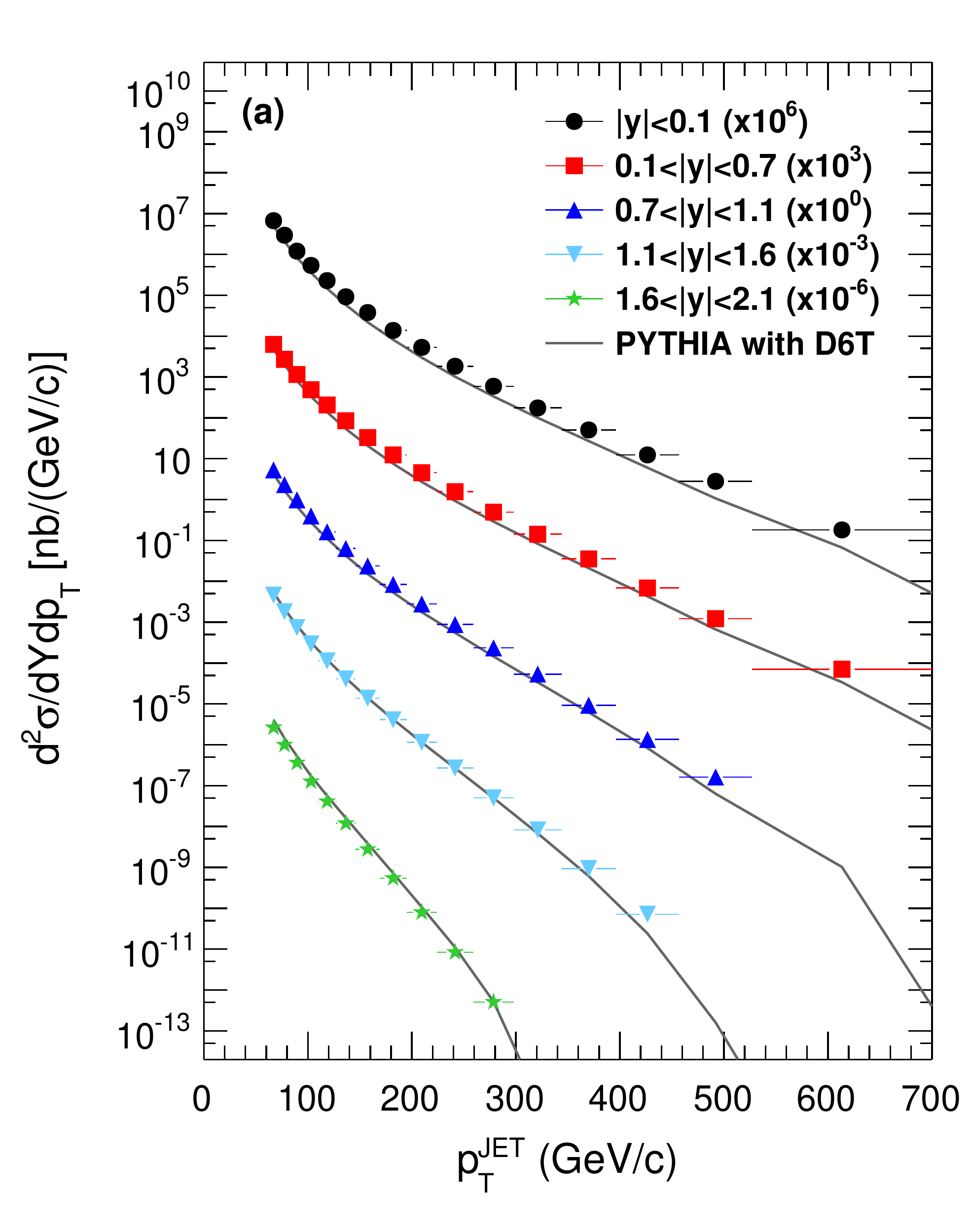}
               \hspace{0.2in}   
    \includegraphics[width=7.7cm]{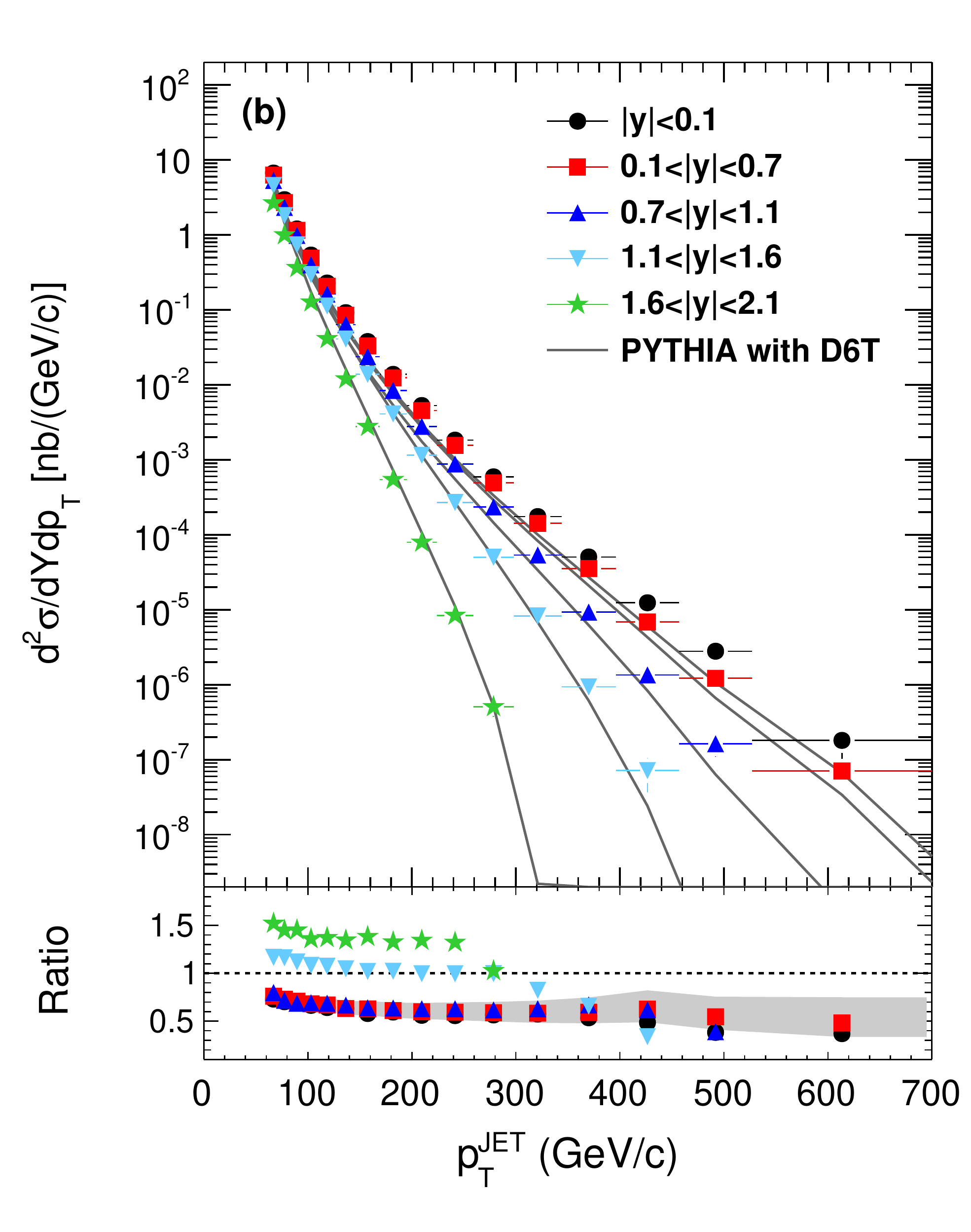}
	\caption{ (a) The inclusive jet cross sections in different rapidity intervals.  CDF values (filled markers) measured at hadron level using the midpoint algorithm (cone radius=0.7); error bar is statistical only.  PYTHIA generated with D6T tune (solid lines) using an iterative cone algorithm (radius=0.7).  An arbitrary factor of $10^{3}$ separates the different rapidity intervals as in the CDF paper for clarity.   
			  (b) The same distributions as in (a), but with all rapidity bins on the same scale. The ratios of the PYTHIA jet cross sections to the measured values are plotted in the lower panel with systematic uncertainties drawn for the measurement in 0.1$<|y|<$0.7 to illustrate the size of the uncertainties involved in the measurement. 
}
	\label{CDFvsPythiaJet}
\end{figure}

Figure~\ref{CDFvsPythiaJet}(a) shows the inclusive jet cross sections from CDF data and PYTHIA, where the different rapidity ($y$) intervals have been scaled by arbitrary factors for clarity of comparison.
In Fig.~\ref{CDFvsPythiaJet}(b), where the different $y$ bins have been plotted on the same scale, it is clear that the jet cross section decreases towards more forward rapidities. In the bottom of the figure, the ratio of the CDF jet cross section to the PYTHIA cross section is shown for each $y$ bin. The inclusive jet cross section obtained from the PYTHIA D6T samples tends to underestimate the measured jet cross section. 
We note, however, that the default $K$-factor of $\sigma_{_{NLO}}/\sigma_{_{LO}}=1$ was used in the generation of our PYTHIA samples.
Therefore, one may claim that the difference would be reduced by applying a greater-than-unity $K$-factor to account for the known difference between $\sigma_{_{NLO}}$ and $\sigma_{_{LO}}$
\footnote{For example, the $K$-factor for the inclusive charged particle spectra at Tevatron energy ($\sqrt{s}=$1.8~TeV) is found phenomenologically to be slightly above unity (1.28)~\cite{Kfactors}}
Other than the underestimation of the overall scale, PYTHIA describes the rapidity dependence of the jet spectra quite well for a wide range of jet-$p_T$. 
This study shows that the inclusive charged particle $p_T$ spectra should not differ by more than a factor of 3 due to differences in the measured and PYTHIA jet cross sections (see lower panel of Fig.~\ref{CDFvsPythiaJet}(b)). 

\section{Convolution with PYTHIA fragmentation functions}
\subsection{\it PYTHIA fragmentation functions}

The fragmentation functions defined in Eq.~\ref{eq3} are obtained from the same PYTHIA samples described in the previous section for the jet cross sections. 
Since these samples were generated in a series of $\hat{p_T}$ bins, some care must be taken to ensure that the contributions from each $\hat{p_T}$ sample are properly accounted for in both the numerator and the denominator of Eq.~\ref{eq3}. 
Specifically, the fragmentation functions are determined via the following sum:

\begin{eqnarray}
F\!F_{i}(p_{T},p^{i}_{T,jet}) & = & \frac{\Delta (\frac{d\sigma_{had}}{dp_{T}dy_{had}}\vert_{|y_{had}|<1.0})}{\Delta \sigma_{jet}\vert_{{|y_{jet}|<y'_{jet}}}}\Bigg\vert_{p_{T,jet} = p^{i}_{T,jet}} \nonumber \\
 & = & \sum_{j} \Bigg[ \frac{\Delta(dN_{had}^{j}/dp_{T}dy_{had})}{N_{event}^{j}/\sigma_{event}^{j}} \Bigg] \Bigg/ \sum_{k} \Bigg[\frac{\Delta N_{jet}^{k}}{N_{event}^{k}/\sigma_{event}^{k}}\Bigg],
\label{eq4}
\end{eqnarray}

\noindent where $N^{j}_{event}$ and $\sigma^{j}_{event}$ are the number of events and the cross section for the $j^{th}$ $\hat{p_T}$ bin, respectively. 
Each fragmentation function is evaluated at $p_{T,jet}=p^{i}_{T,jet}$ with a width of $\Delta p^{i}_{T,jet}$ for the $i^{th}$ jet-$p_T$ bin. 
$N_{had}$ and $N_{jet}$ are the number of charged particles and the number of jets in $|y_{had}|<$ 1.0 and $|y_{jet}|<y'_{jet}$, respectively.  (Note that $y_{had}$ and $y_{jet}$ are not necessarily the same.) 

\begin{figure}[htbp]
	\centering
    \includegraphics[width=7.3cm]{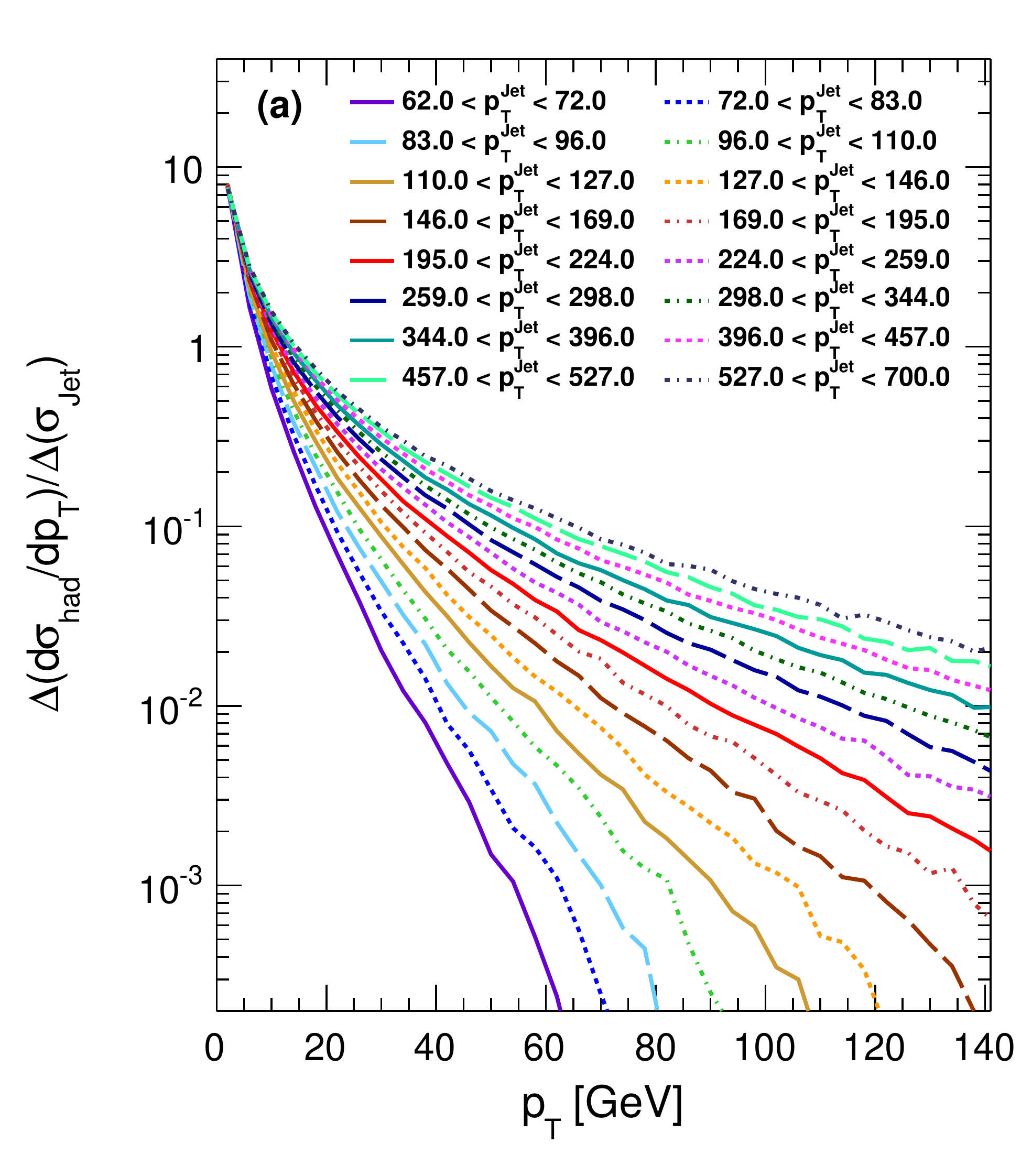}
       \hspace{0.2in}
   \includegraphics[width=7.3cm]{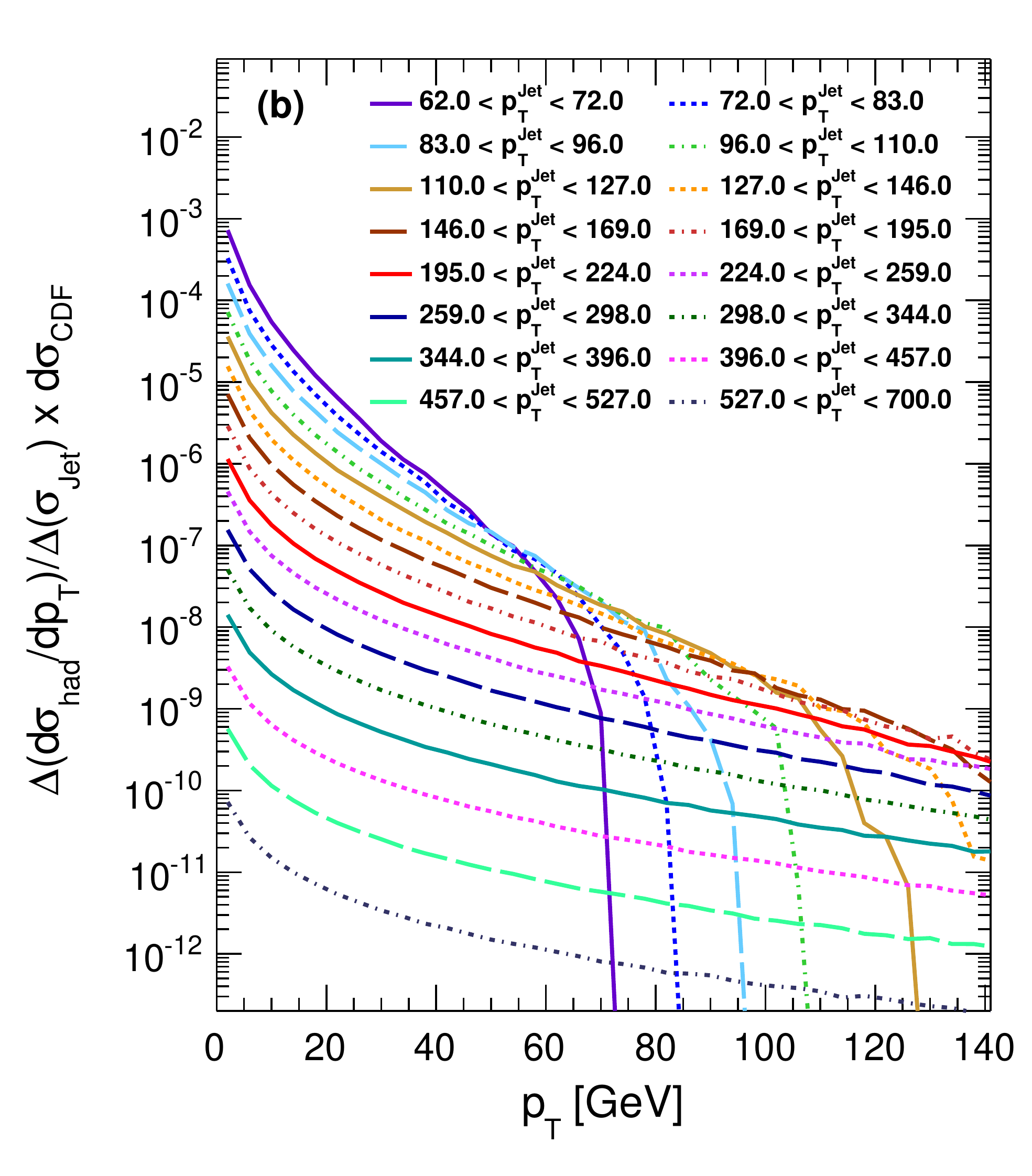}
	\caption{ (a) PYTHIA D6T fragmentation function in each jet-$p_T$ bin. (b) The same fragmentation functions weighted by the associated jet cross section measured by CDF. $d\sigma_{_{CDF}}$ corresponds to $d\sigma_{jet}/dp_{T}dy$ in Eq.~\ref{eq2_1}.}
	\label{PythiaFFAll}
\end{figure}

The PYTHIA fragmentation functions $F\!F_{i}(p_{T},p^{i}_{T,jet})$ for $|y_{had}|<1.0$ and $|y_{jet}|<1.0$ are shown in Fig.~\ref{PythiaFFAll}(a) for $p_T<140$~GeV/c and bins of $p^{i}_{T,jet}$ corresponding to the CDF jet cross-section measurement. 
The hardening of the fragmentation function with increasing jet-$p_T$ is apparent. 
The same fragmentation functions are shown in Fig.~\ref{PythiaFFAll}(b), where each has been weighted by the associated jet cross section in order to compare their relative contributions to the charged particle spectra. 
Due to the steeply falling nature of the jet cross section, the vast majority of charged particles in the region of interest (30 GeV/c $<p_T<$ 140 GeV/c) are from jets with $p_T < 400$~GeV/c. 
The contribution from higher $p_T$ jets is less than 0.1$\%$. 

Although a direct verification of the PYTHIA fragmentation functions in Fig.~\ref{PythiaFFAll} is not possible in the absence of the same measurement on 1.96 TeV data, reasonable (and sometime very good) agreement
\footnote{There is some indication that the fragmentation properties of quark jets are rather poorly described by PYTHIA when quark and gluon jet are investigated separately~\cite{JetFF}.}
has been seen between PYTHIA and a variety of fragmentation-related measurements~\cite{JetFF, JetFF2, JetFF3, CDFJetShape}. 
For example, detailed CDF studies of inclusive jet shapes at $\sqrt{s}$=1.96 TeV~\cite{CDFJetShape} show that the integrated and differential jet shapes are well described by PYTHIA Tune A for jet-$p_T$ up to 380 GeV/c. 
Furthermore, the fact that PYTHIA and NLO charged particle spectra are in reasonable agreement~\cite{FA_DD_AY}, where the latter uses fragmentation functions  based on global data fits (AKK~\cite{AKK}, DSS~\cite{DSS}, HKNS~\cite{HKNS}), implies that PYTHIA fragmentation should be comparable to these global fits. 
In this case, the fragmentation functions obtained from PYTHIA should be a reliable proxy for the global understanding of fragmentation from data. 
Regardless, this paper investigates discrepancies significantly larger than 50\%, so the detailed matching of the PYTHIA fragmentation model to measurement is not our primary concern. 
Moreover, model-independent fragmentation functions will be introduced later in Section 5. 

\subsection{\it Convolution of PYTHIA jet spectrum with PYTHIA fragmentation functions}

\begin{figure}[htbp]
	\centering
	\includegraphics[width=7.7cm]{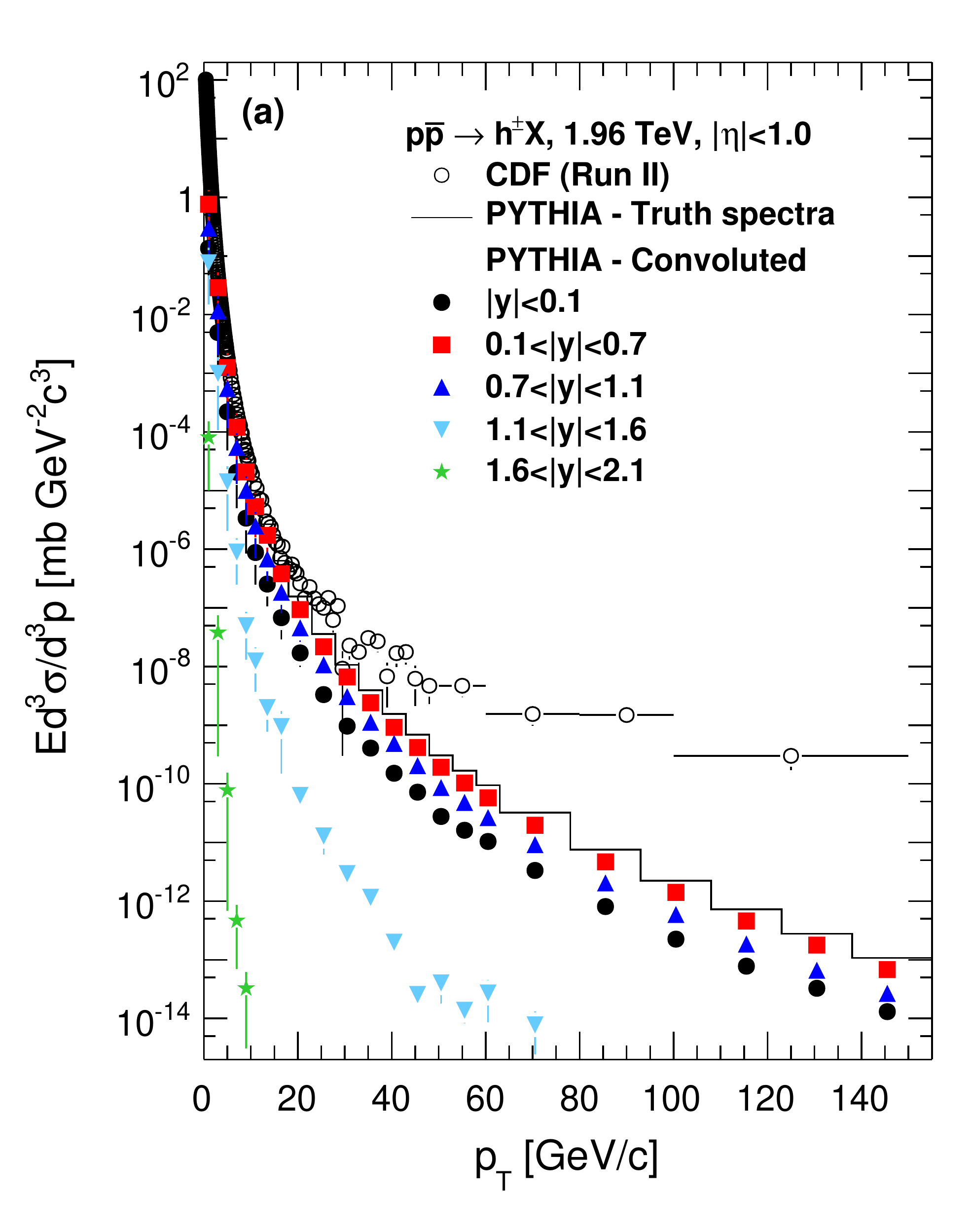}
     \hspace{0.2in}
	\includegraphics[width=7.7cm]{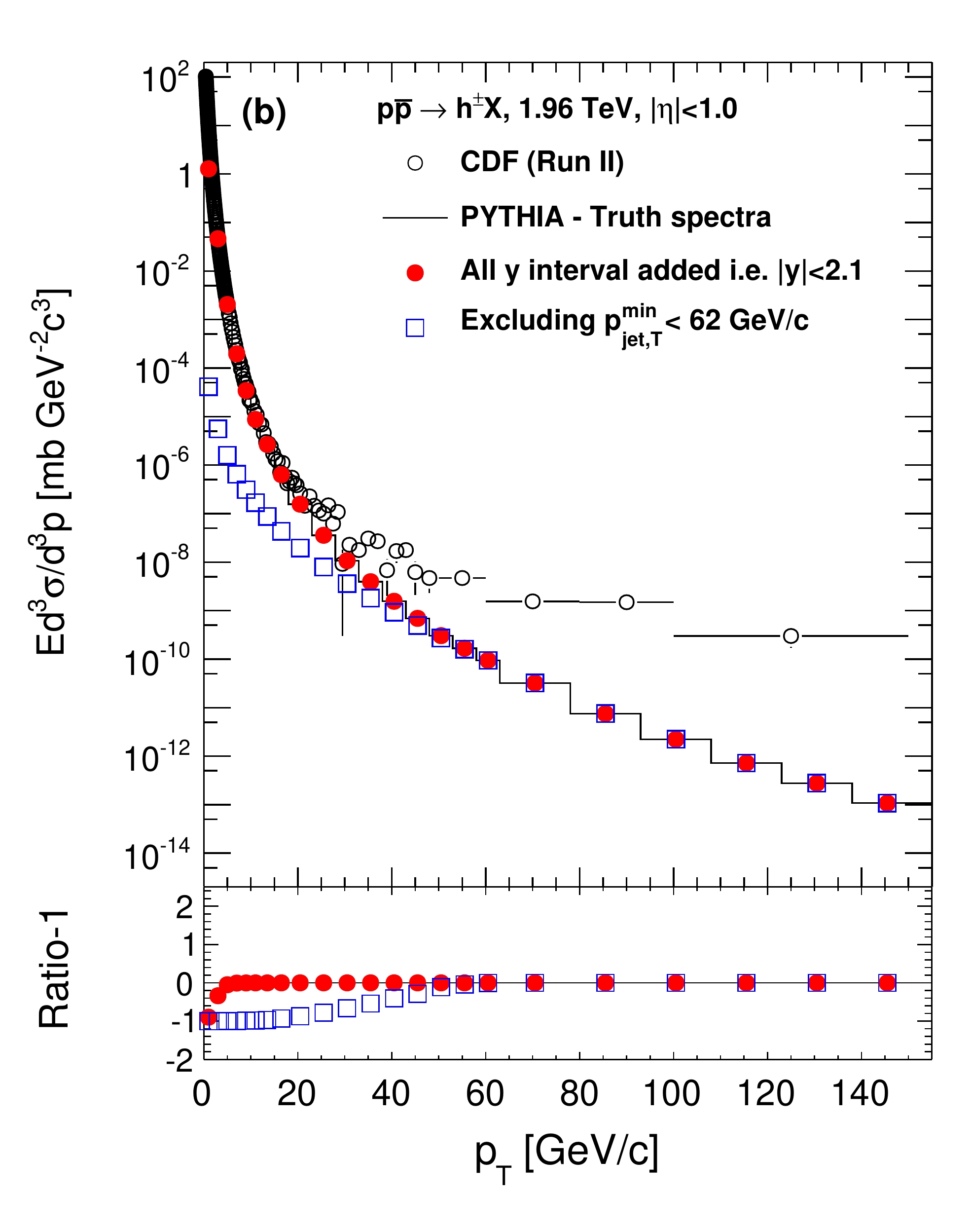}
	\caption{(a) PYTHIA charged particle cross section obtained from the convolution of the PYTHIA inclusive jet cross sections in different jet $y$ ranges with PYTHIA fragmentation functions (filled markers) compared to the ``true" charged particle cross section from PYTHIA (solid line).  (b) The convoluted spectra after summing the contributions from all jet $y$ ranges (filled circles), and the same after excluding the contribution from jets with $p_T<62$~GeV/c (open squares). In both figures, the CDF measured cross sections are also shown (empty circles) for comparison.}
	\label{ConvolutedPythiaPythia_v1}
\end{figure}

To test the effectiveness of the convolution technique put forth in Section~\ref{sec:method}, we first attempt to retrieve the known PYTHIA charged particle $p_T$ spectrum from a convolution of the PYTHIA jet spectra with the PYTHIA fragmentation functions calculated in Eq.~\ref{eq4} and plotted in Fig.~\ref{PythiaFFAll}. 
We perform the convolution separately for each of the five jet rapidity intervals (see Fig.~\ref{CDFvsPythiaJet}), in order to quantify their relative contributions to the single particle spectra.  
The results of this test are shown in Fig.~\ref{ConvolutedPythiaPythia_v1}, where the PYTHIA charged particle spectrum (solid line) is compared to the output of the convolution for charged particles with $|\eta|<1.0$.  
The relative contribution of jets in different $y$ ranges to $|\eta|<1.0$ charged particles can be seen in Fig.~\ref{ConvolutedPythiaPythia_v1}(a), where the rapidity ranges are indicated by the same symbols as in Fig.~\ref{CDFvsPythiaJet}. 
As expected from the measured jet shapes\footnote{From measured jet shapes, we know that the majority of the energy in a jet is concentrated around jet axis.  So, the rapidity of the leading tracks should correlate quite closely with the rapidity of the jet.}~\cite{CDFJetShape}, the large majority of particles fragment from jets within $|y|<1.1$.  
In Fig.~\ref{ConvolutedPythiaPythia_v1}(b), the contributions from all rapidity intervals are summed and compared to the true charged particle spectrum. 
In the bottom of that figure, the ratio of the convoluted spectra to the true charged particle spectra is shown. Except at low $p_T$ (below a few GeV/c) where non-perturbative particle production and jet reconstruction inefficiency become relevant, the convolution method reproduces the true spectra almost exactly (well within 0.01$\%$ for $p_T\geq$ 6 GeV/c).

\subsection{\it Convolution of measured jet spectrum with PYTHIA fragmentation functions}

With the robustness of the convolution technique verified on PYTHIA, the next step is to introduce the CDF measurement of the jet spectrum in the convolution.
Given the level of agreement already demonstrated in Section~\ref{sec:jetspectra} between PYTHIA and the measured jet spectra, one should expect the convolution based on the measured jet spectra to result in quite similar charged particle spectra as already seen for PYTHIA in Fig.~\ref{ConvolutedPythiaPythia_v1}. 
However, there is an additional complication involved in using the measured jet spectra in the convolution instead of the generated PYTHIA events. 

Unlike for PYTHIA where the jet cross section is available down to very low $p_T$, the jet cross section measured by CDF has only been published between 62 and 700 GeV/c. 
The effect on the charged particle spectrum from excluding the fragmentation products of jets below 62~GeV/c is shown for PYTHIA in Fig.~\ref{ConvolutedPythiaPythia_v1}(b) represented by empty squares. 
These low-$p_T$ jets contribute significantly to the charged particle spectrum up to around 40 GeV/c. 
As expected, the contribution completely vanishes at $p_T=62$~GeV/c, since a jet cannot fragment into a more energetic charged particle.
Since the disagreement with (N)LO pQCD calculations is most prominent above $p_T$ $\geq$ 50 GeV/c, a detailed understanding of the low-$p_T$ contributions is not central to this investigation.  
Hereafter, whenever the measured jet cross section is used in the convolution, we use the PYTHIA value for $p_T<$~62 GeV/c. 

\begin{figure}[htbp]
	\centering
    \includegraphics[width=7.7cm]{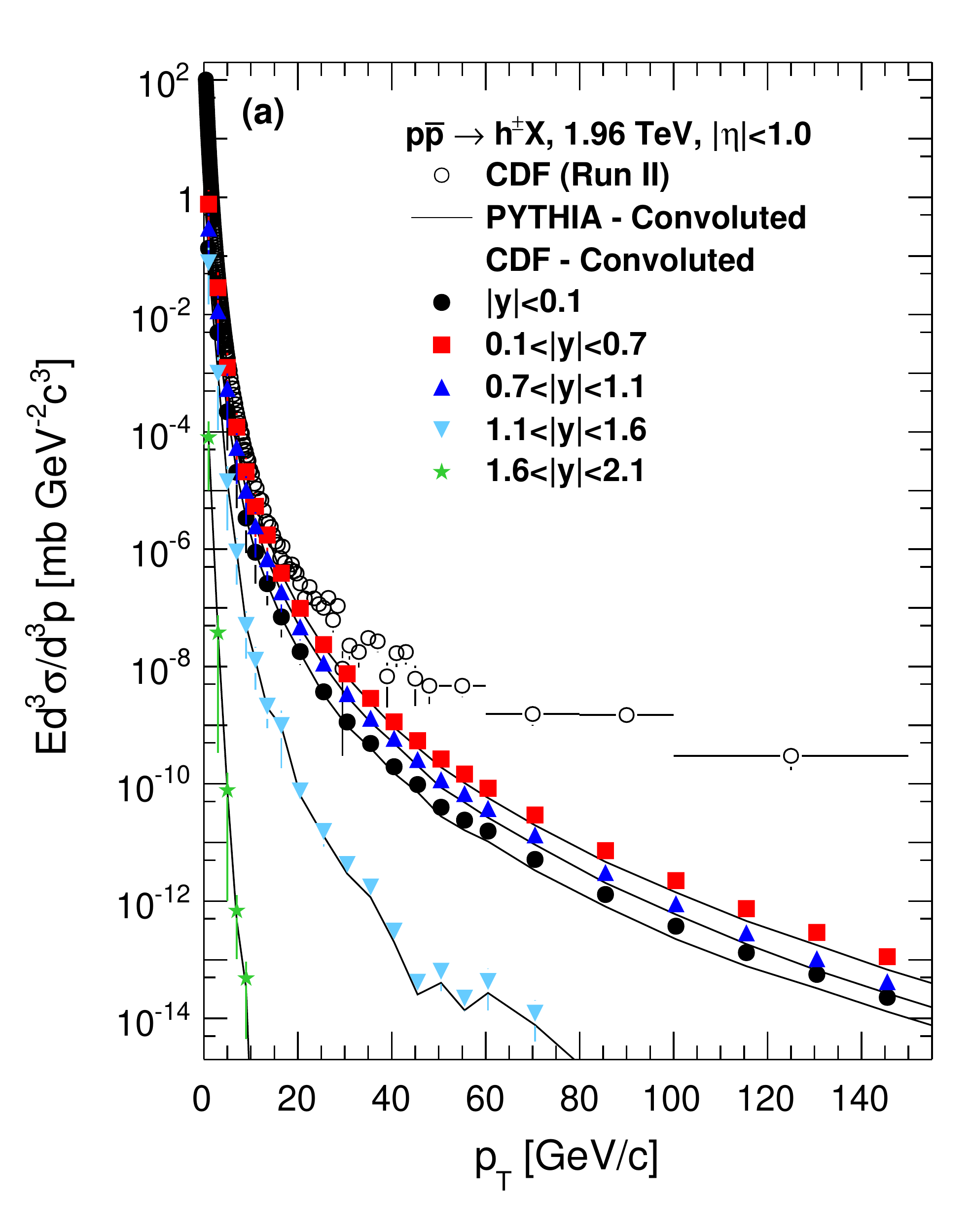}
       \hspace{0.2in}
  \includegraphics[width=7.7cm]{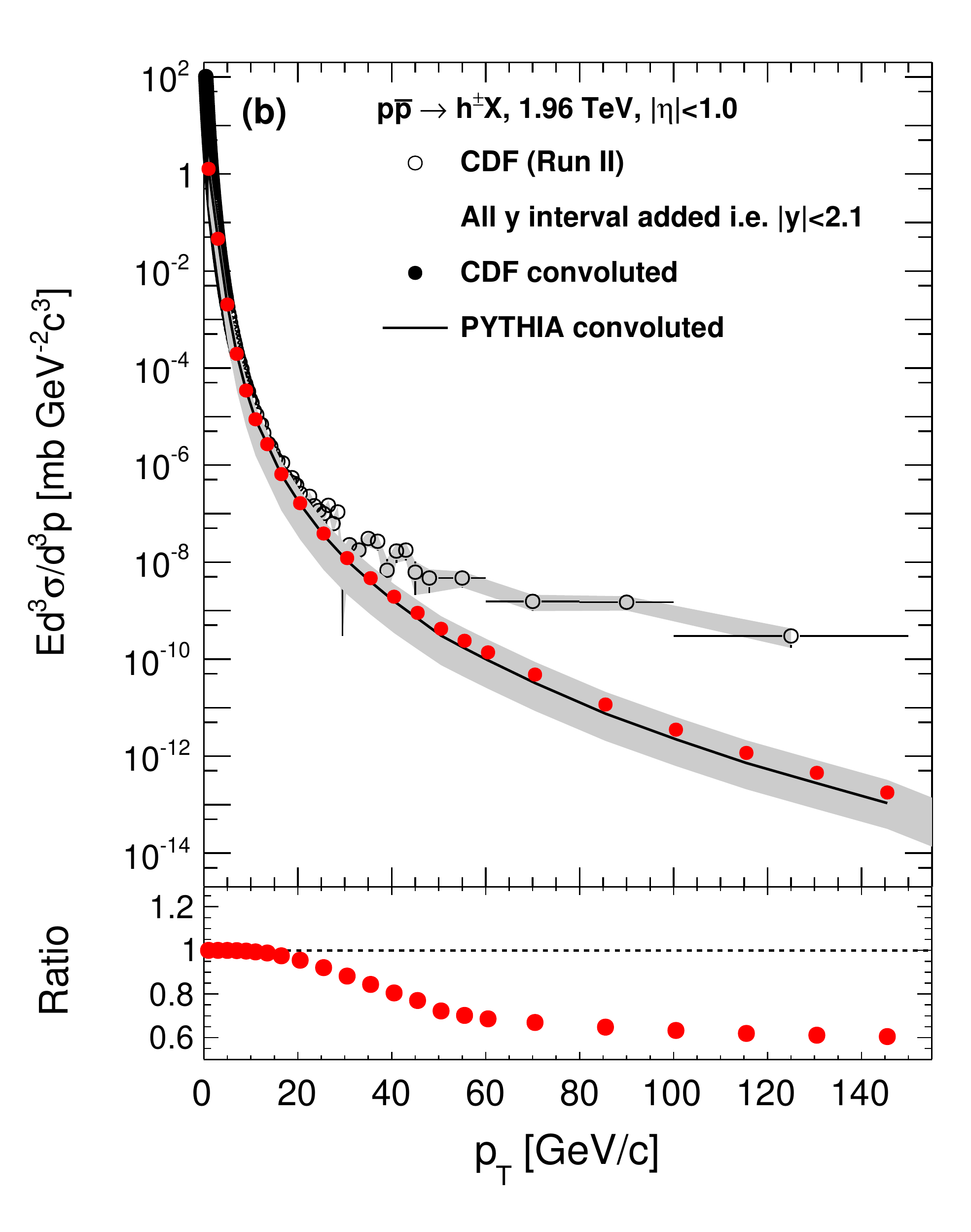}	
	\caption{(a) Charged particle $p_T$ differential cross section obtained from the convolution of the CDF jet cross sections measured in five rapidity intervals with the PYTHIA fragmentation functions (filled markers). Same cross sections obtained from the PYTHIA-only convolution are also shown in comparison (black line). (b) The convoluted spectra after summing the contributions from all jet $y$ ranges (filled circle), and the same from the PYTHIA-only convolution (black line).  The ratio of the two resulting charged particle spectra (from convolution with either the CDF or PYTHIA jet spectra) is shown in the lower panel.   }
	\label{ConvolutedPythiaCDF_v1}
\end{figure}

In Fig.~\ref{ConvolutedPythiaCDF_v1}(a), the charged particle $p_T$ spectra are shown for the convolutions based on the measured jet cross sections, again using the same symbol conventions for the rapidity ranges as in Fig.~\ref{CDFvsPythiaJet}. 
In Fig.~\ref{ConvolutedPythiaCDF_v1}(b), the contributions from all rapidity intervals are summed as it is done in~\ref{ConvolutedPythiaPythia_v1}(b). 
To see the variation of the obtained spectrum due to the uncertainties in the CDF jet measurement, a conservative choice of $82\%$, which is the largest systematic uncertainty from the measurements in 
the first three rapidity intervals for jet-$p_T$ below 457 GeV is applied to the obtained charged particle spectrum, which is shown as a grey band. Also, a grey band is drawn for the measured charged particle spectrum to indicate the size of the uncertainties in the measurement~\footnote{Statistical uncertainty only, although the statistical uncertainties are comparable to the total uncertainties for $p_T \geq 50$~GeV/c~\cite{CDFsingle}}. The spectra from the PYTHIA-only convolution is drawn as black lines for comparison in both~\ref{ConvolutedPythiaCDF_v1}(a) and~\ref{ConvolutedPythiaCDF_v1}(b).
In the bottom of the figure, the ratio of the resulting spectrum from the convolution of PYTHIA jet spectra to that of CDF jet spectra is shown. As expected from the agreement between the CDF measured jet spectra and PYTHIA (better than $\pm50\%$ below 400~GeV), the convoluted charged particle spectra are in similar agreement. 

\section{Convolution with toy-model fragmentation functions}

Knowing that the charged particle spectra from the convolution of the measured jet spectra with PYTHIA fragmentation functions still vastly undershoot the measured spectrum, it is interesting to know how sensitive the charged particle spectra are to arbitrary changes in the fragmentation function; or rather, how large a change in the fragmentation functions would be required to reproduce the measured spectrum?
We address this question by convoluting the measured jet spectra with the fragmentation functions obtained based on the following two toy fragmentation models:
\begin{enumerate}
\item{Harder fragmentation -- modified shapes}
\item{Hardest possible fragmentation -- one charged particle per jet}
\end{enumerate}
The modified fragmentation functions are obtained in the same manner as for the default PYTHIA (see Eq.~\ref{eq2_1}). 

\subsection{\it Harder fragmentation}

We make the default PYTHIA fragmentation functions harder by modifying the shape of the fragmentation functions. 
Of course, there are arbitrarily many imaginable functional forms one could use for ``hardening'' the fragmentation functions. 
Here we have chosen to harden the fragmentation functions by bending the $p_{T}$-shape with a simple power-law functional weight:
\begin{equation}
f(p_T) = 1 + c(p_T - p_{T}^{min})^{n}  \mbox{~~for~~} p_T > p_{T}^{min} ,
\label{harderFF}
\end{equation}

\begin{figure}[htbp]
	\centering
   \includegraphics[width=7.3cm]{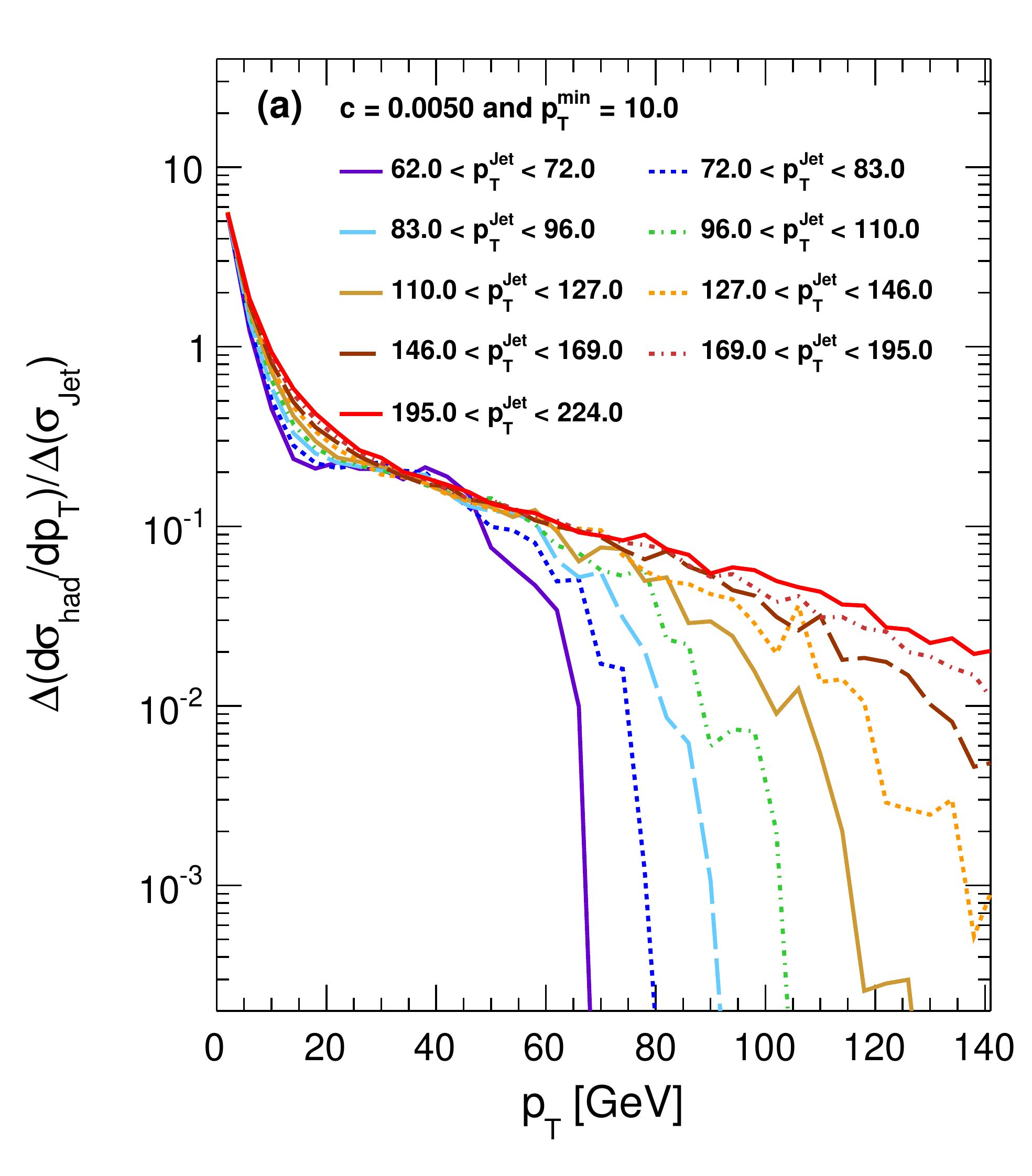}
     \hspace{0.3in}
  \includegraphics[width=7.3cm]{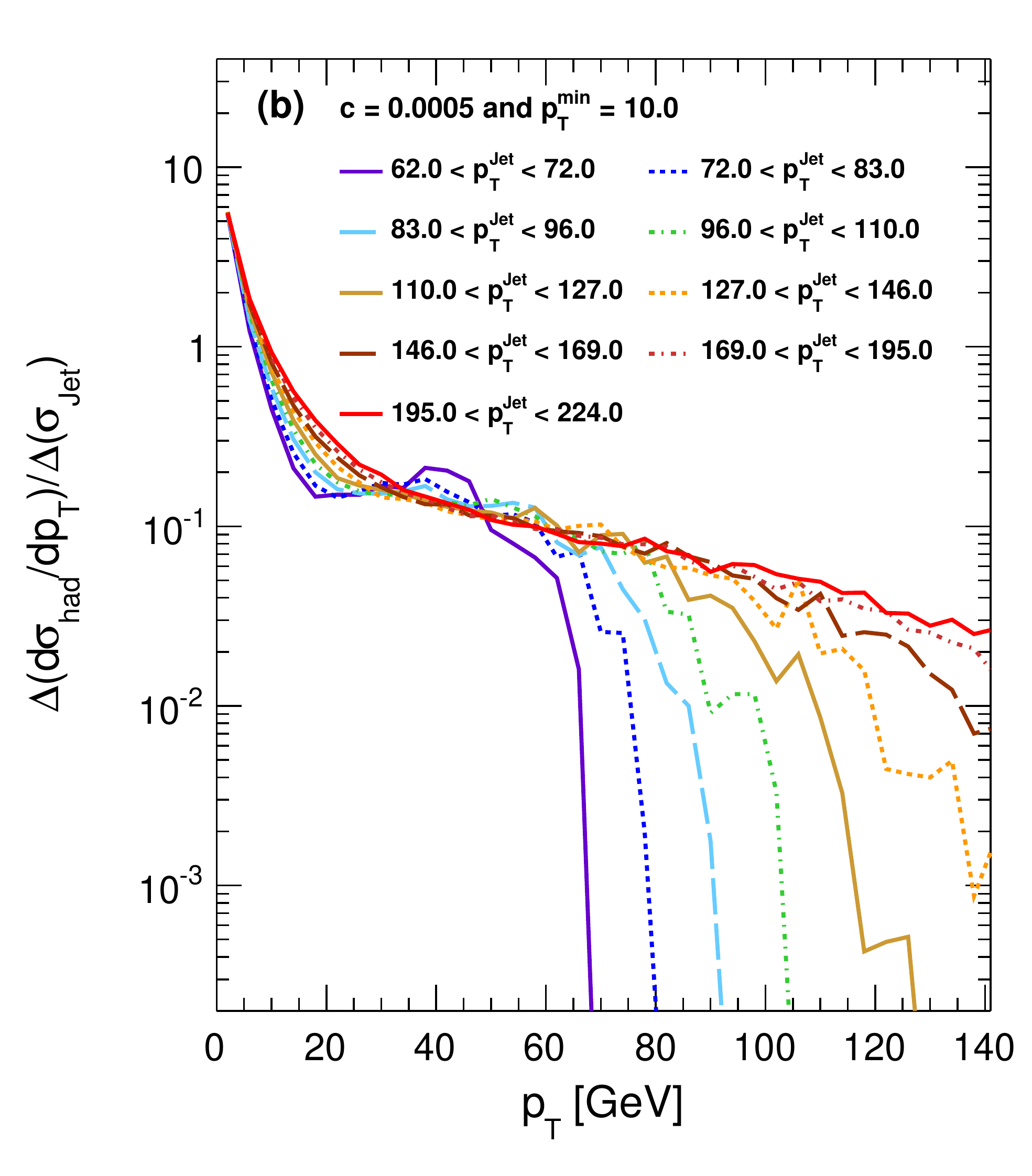}
 \includegraphics[width=7.3cm]{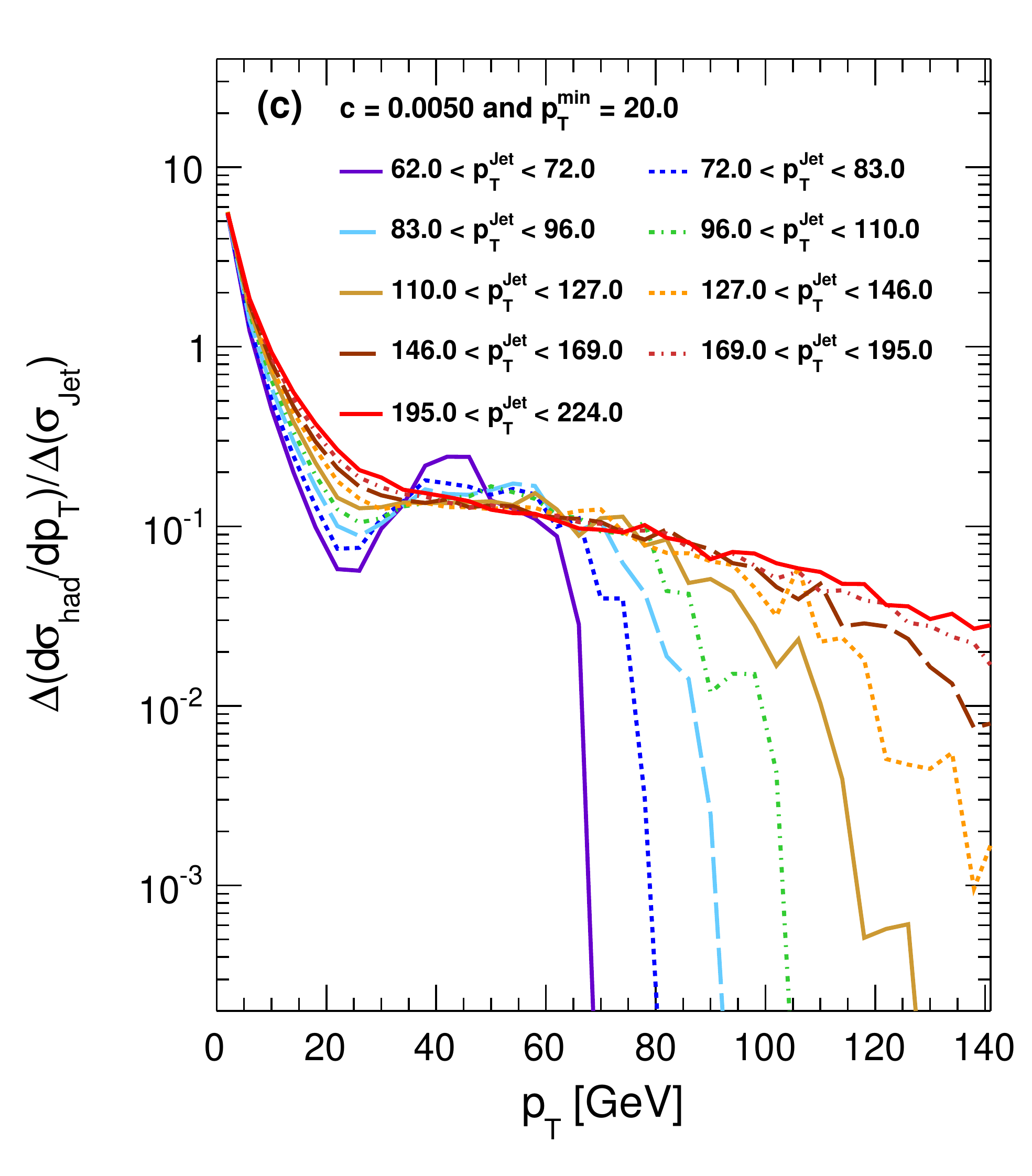}
      \hspace{0.3in}
   \includegraphics[width=7.3cm]{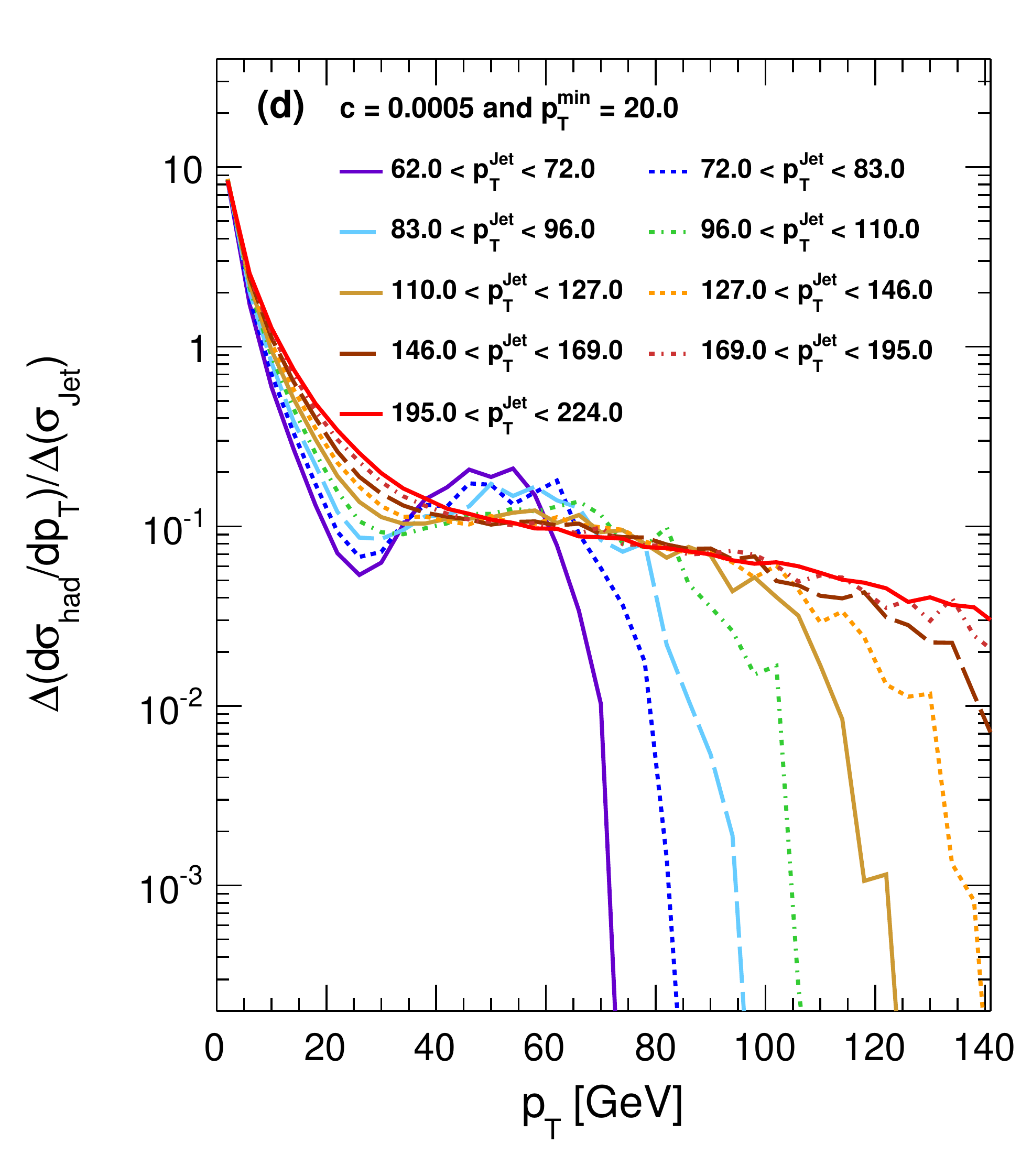}
	\caption{(a-d) PYTHIA (with D6T) fragmentation function in each jet-$p_T$ bin modified by the power-law function (see Eq.~\ref{harderFF}) for different choices of the ``hardening parameters", $c$ and $p_{T}^{min}$.  The same color and line style conventions are used from Fig.~\ref{PythiaFFAll}.  }
	\label{PythiaModFF2}
\end{figure}

\noindent The shape of the modification is governed by the constant $c$ and the exponent $n$, while $p_{T}^{min}$ determines the $p_{T}$ value where the modification begins.
To maximize the effect in this study, we bend the shape of the fragmentation function to the extent that the energy sum of the fragmented charged particles equals  the energy of the corresponding jet. 
In this fashion, the fragmentation function can be maximally hardened for any choice of two parameters: $c$ and $p_{T}^{min}$. 
We try four different combinations of $c$ and $p_{T}^{min}$, which are presented in Fig.~\ref{PythiaModFF2}.
The values of $c$ (0.005 and 0.0005) were selected to ensure the bending is not unrealistically abrupt.
Since the spectrum from the convolution already describes the CDF measurement rather well up to 20 GeV/c, we pick $p_{T}^{min}$ accordingly (10 and 20 GeV/c) to maintain this agreement. 
With these choices of $c$ and $p_{T}^{min}$, the exponent $n$ that fulfills the maximum hardening ranges from 1 to 4 for the various jet-$p_T$ bins. 
The modified fragmentation functions from jets within a range of $0.1<|y|<0.7$ are shown in Fig.~\ref{PythiaModFF2} for the different values of $c$ and $p_{T}^{min}$. 
Here, we only show the fragmentation functions for jets with $p_T$ up to 400 GeV/c, as the fragmentation functions with higher jet $p_T$ have a negligible contribution to the inclusive spectrum (see Fig.~\ref{PythiaFFAll}(b)). 
When compared to Fig.~\ref{PythiaFFAll}(a), the hardening of the fragmentation functions is immediately apparent. 
Despite an attempt to maintain smooth functions, the fragmentation functions in the first few jet-$p_T$ bins are if anything unrealistically hard, as they contain regions with positive slopes. 

For each of the four scenarios, the convolution of the hardened fragmentation functions with the CDF jet measurement is performed for the three dominant rapidity intervals ($|y|<$0.1, 0.1$<|y|<$0.7, 0.7$<|y|<$1.1), which are then summed.
In Fig~\ref{convl_cdf_mod2ff}(a), the resulting charged particle spectra are compared to the CDF measured charged particle spectrum. The harder fragmentation functions are reflected in a hardening of the inclusive spectrum. 
The disagreement between the measurement and the resulting spectra is less pronounced between 30 and 60 GeV/c. 
However, a disagreement of up to two orders of magnitude remains at the highest measured transverse momentum. 

\subsection{\it Hardest possible fragmentation}
Finally, we assume the extreme case of hard fragmentation, for which each hard-scattered parton fragments into a single charged particle.
This is the hardest possible fragmentation ($z=1$) that still conserves energy (though not necessarily other conserved quantities). 
Unlike the previous fragmentation functions, which are still model-dependent to some extent, the $z=1$ scenario is model-independent. 

\begin{figure}[htbp]
	\centering
    \includegraphics[width=7.7cm]{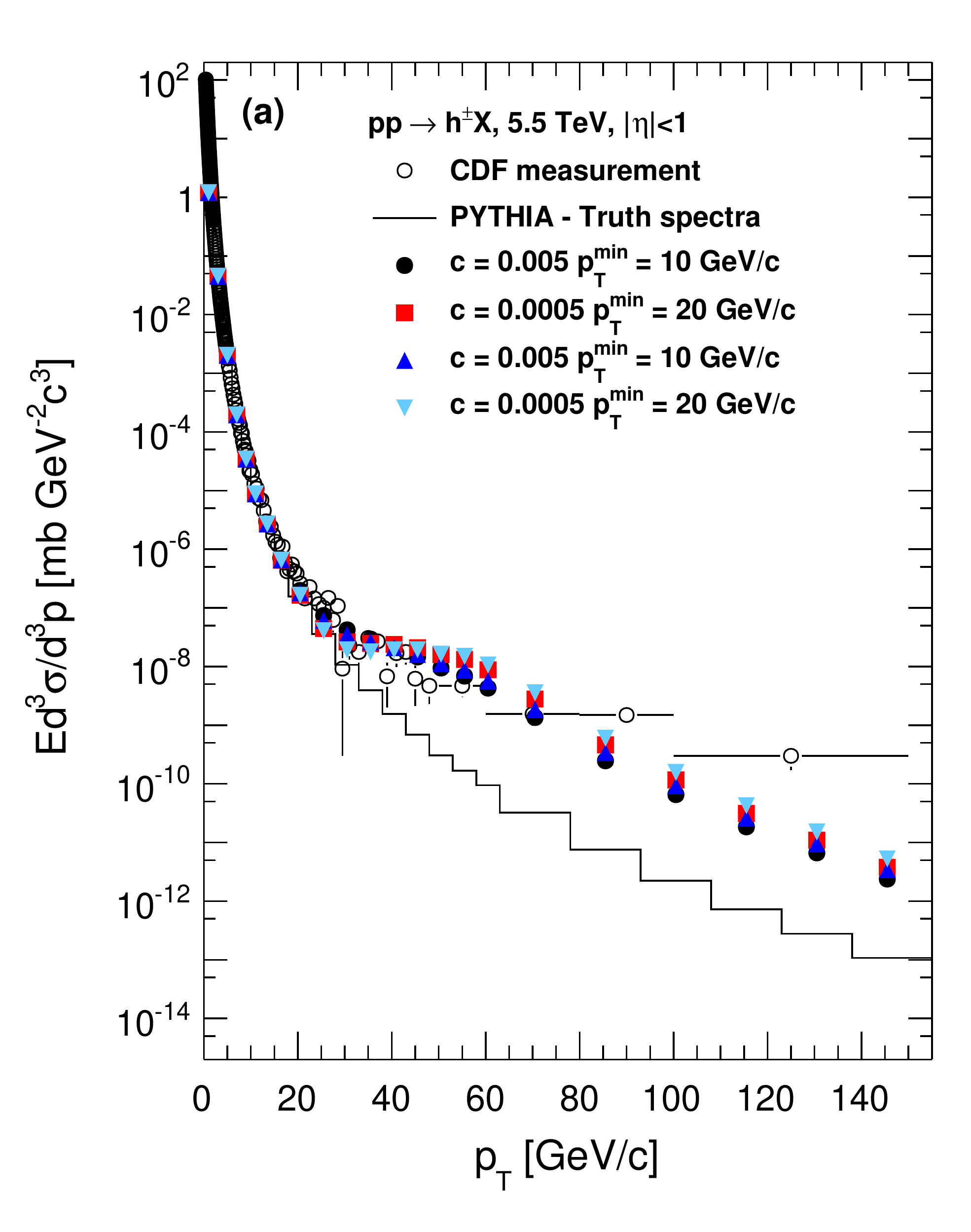}
    \hspace{0.2in}
     \includegraphics[width=7.7cm]{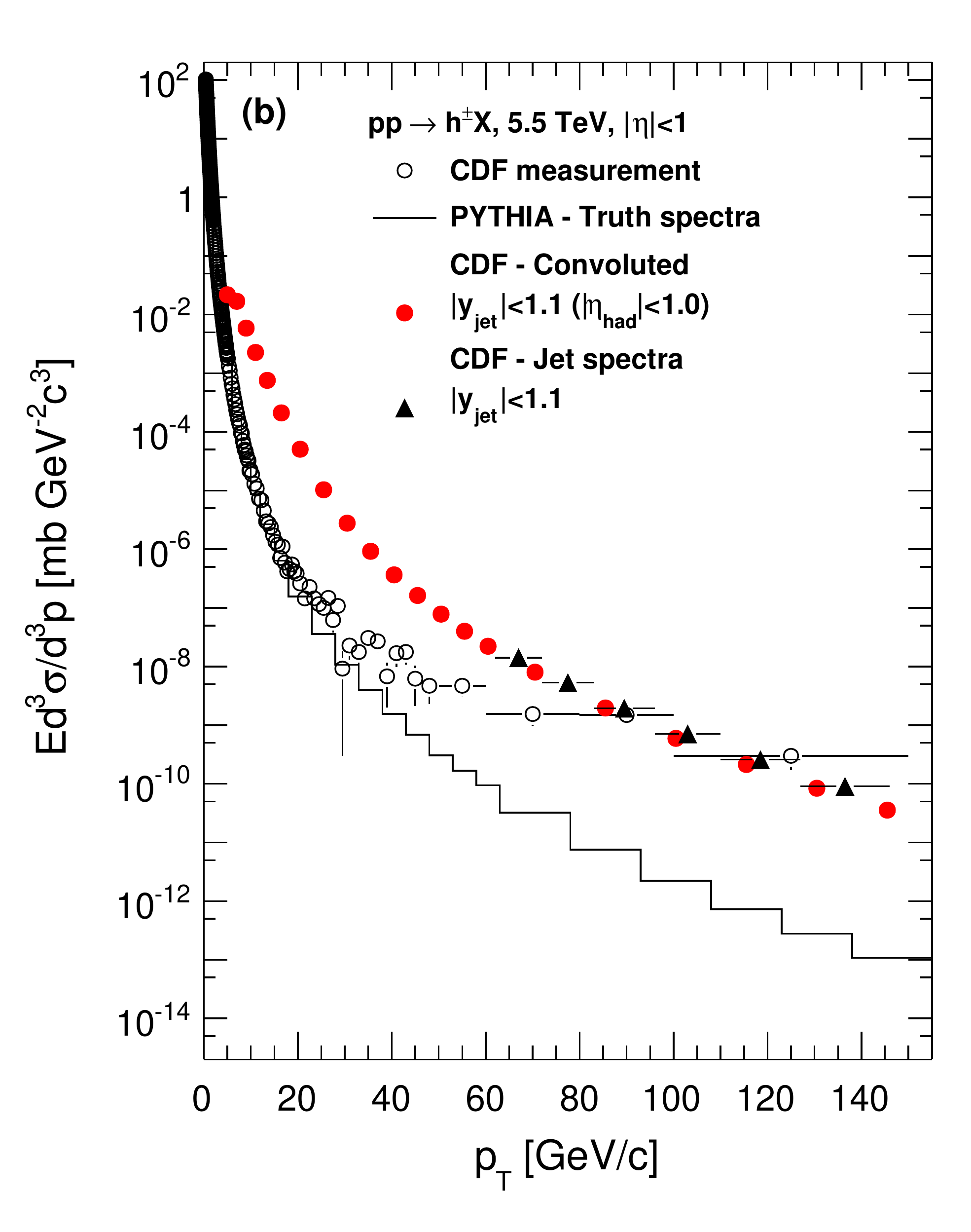}
	\caption{(a) Charged particle $p_T$ differential cross sections obtained from the convolution of the measured jet cross sections with PYTHIA fragmentation functions modified according to Eq.~\ref{harderFF}. Different symbols show the spectra obtained with the modified fragmentation functions based on different sets of hardening parameters. (b) The same cross sections obtained from the convolution of measured jet cross sections ($|y|<$1.1) with the hardest possible fragmentation (filled circles) compared to PYTHIA (solid line).  The CDF combined jet spectra averaged over $|y|<$1.1 (filled triangle)  and the CDF charged particle spectrum (open circle) are plotted for comparison.}
	\label{convl_cdf_mod2ff}
\end{figure}

In Fig ~\ref{convl_cdf_mod2ff}(b), the result of applying the $z=1$ fragmentation to the measured jet spectrum is shown. 
Again, the contributions from the three dominant rapidity intervals are summed to obtain the inclusive charged particle spectrum for $|\eta|<$1. 
Both the shape as well as the overall magnitude are dramatically changed, enhancing the cross section by a few orders of magnitude. In the same figure, the jet spectra for the three $y$ intervals ($|y|<$0.1, 0.1$<|y|<$0.7, 0.7$<|y|<$1.1) are averaged and plotted in the form of an invariant yield for comparison. However, the charged particle spectrum measured by CDF at the highest $p_T$ actually exceeds their measured jet spectra with $z=1$ fragmentation.  Furthermore, the level of underestimation is beyond the level of systematic uncertainties shown in the CDF jet measurement, which range from 10 to 80$\%$ depending on jet-$p_T$~\cite{CDFjet1}.

\section{Summary and discussion}

The latest CDF measurement of the inclusive charged particle $p_T$ spectrum up to very high $p_{T}$ is cross-checked against the convolution of the measured inclusive jet spectra with a set of different fragmentation functions. 
The inclusive charged particle spectrum obtained from the PYTHIA fragmentation functions convoluted with the measured jet cross section fails to reproduce the measured spectrum, despite matching (N)LO predictions reasonably well. 
PYTHIA fragmentation functions modified by a simple toy model result in quite different spectra shapes.
However, despite an improved agreement at intermediate $p_T$, the convolutions still underestimate the measured charged particle spectrum at high $p_T$. 
Finally, we show that even the most extreme case of each jet fragmenting into a single charged particle fails to reconcile the measured jet and charged particle spectrum.
Based on these studies, we rule out the possibility that the disagreement of the NLO pQCD calculations with the measured charged particle spectrum (but not the measured jet spectra) is due to an unexpectedly hard fragmentation of jets. 
Thus, we conclude that the CDF charged particle spectrum cannot be reconciled with the present understanding of factorized pQCD \footnote{This doesn't rule out the possibility discussed in ~\cite{AKK} that the factorization breaks down, since this investigation with the convolution method relies upon the validity of the factorization theorem.}. 

This work was originally motivated for the purpose of cross-checking the CDF charged particle $p_T$ spectrum measurement, but this convolution technique can be extended for use in future measurements without relying on input from PYTHIA at all.
Any of the following three measurements can be independently obtained from the (de)convolution of the other two: the inclusive jet spectrum, the fragmentation functions of inclusive jets, and the inclusive charged particle spectrum.
We expect that such a cross-check of future single particle $p_T$ spectra measurements against the available calorimetric information will be very helpful for understanding the tracking-related systematic uncertainties in the high-$p_T$ regime.
\\
\\
{\it Note added}: Another analysis reaching similar conclusions~\cite{NoteOnCDF} was submitted for publication during the finalization of this paper. 

\section{Acknowledgment }
ASY thanks David d'Enterria and Francois Arleo for discussions.


\end{document}